\documentclass[fleqn,12pt,twoside]{article}
\usepackage[headings]{espcrc1}

\readRCS
$Id: espcrc1.tex,v 1.2 2004/02/24 11:22:11 spepping Exp $
\usepackage{graphicx}
\usepackage[figuresright]{rotating}

\newcommand{\AmS}{{\protect\the\textfont2
  A\kern-.1667em\lower.5ex\hbox{M}\kern-.125emS}}

\title{Deformation Effects in Hot Rotating $^{46}$Ti 
Probed by the Charged Particle Emission 
and GDR $\gamma$-Decay}

\author{M.~Brekiesz\address[IFJ]{The H. Niewodnicza\'nski Institute of Nuclear Physics, 
        PAN, 31-342 Krak\'ow, Poland}, %
        A.~Maj\addressmark,
        M.~Kmiecik\addressmark.
        K.~Mazurek\addressmark,
        W.~M\c{e}czy\'nski\addressmark,
        J.~Stycze\'n\addressmark,
        K.~Zuber\addressmark,
        P.~Papka\address[IPHC]{IPHC and ULP (Strasbourg I), B.P. 28 F-67037 Strasbourg Cedex 2, France}\address[iThemba]{iThemba LABS, 7129 Somerset West, South Africa },
        C.~Beck\addressmark[IPHC],
        F.~Haas\addressmark[IPHC],
        V.~Rauch\addressmark[IPHC],
        M.~Rousseau\addressmark[IPHC],
        A.~S\`anchez~i~Zafra\addressmark[IPHC],
        J.~Dudek\addressmark[IPHC]
        ~and
        N. Schunck\addressmark[IFJ]\address[UM]{Depart\'amento de F\'{\i}sica Te\'orica, Universidad Aut\'onoma de Madrid, Spain }}
       
\runtitle{Deformation Effects in Hot Rotating $^{46}$Ti... }
\runauthor{M. Brekiesz et al.}

\begin{document}

% typeset front matter
\maketitle

\begin{abstract}
The $^{46}$Ti$^{*}$ compound nucleus, as populated by the fusion-evaporation
reaction $^{27}$Al + $^{19}$F at the bombarding energy of 
$E_{lab}$ = 144~MeV, has been
investigated by charged particle spectroscopy using the multidetector array
{\sc ICARE} at the {\sc VIVITRON} tandem facility of the IReS (Strasbourg).
The light charged particles and high-energy $\gamma$-rays from the GDR decay
have been measured in coincidence
with selected evaporation residues. The {\sc CACARIZO} code, a Monte Carlo
implementation of the statistical-model code {\sc CASCADE}, has been used
to calculate the spectral shapes of evaporated $\alpha$-particles which are
compared with the experimental coincident spectra. This comparison 
indicates the 
signature of large deformations (possibly superdeformed and hyperdeformed
shapes) present in the compound nucleus decay. 
The occurrence of the Jacobi shape transition is also discussed in the
framework of a newly developed rotating liquid drop model.
\end{abstract}

\section{INTRODUCTION}

In the recent years, there has been a large number of both experimental 
and theoretical
studies aimed at understanding the effects of large deformations in the case of
light-mass nuclei~\cite{Kic03,Maj01,Maj02,Pom06,Dub07,Pap04b,Pap03,Pap04a,Ide,Beck04,Sven,Brekiesz_Zakopane}.
 The very elongated prolate or triaxial shapes were deduced
from the spectra of the Giant Dipole Resonance (GDR) from the decay of $^{45}$Sc$^{*}$ ~\cite{Kic03}
 and $^{46}$Ti$^{*}$~\cite{Maj01,Maj02}.
 The results were consistent with predictions made within the
macroscopic-microscopic approach of the LSD (Lublin-Strasbourg Drop)
model~\cite{Pom06,Dub07}, in which the large deformations are ascribed to the
Jacobi shape transition. The large deformations were also indicated by the
measurement of energy spectra and angular distributions of the 
light charged particles (LCP) originated from the decay 
of $^{44}$Ti$^{*}$ \cite{Pap04b} as formed in two fusion reactions
$^{16}$O + $^{28}$Si~\cite{Pap03} and $^{32}$S + $^{12}$C~\cite{Pap04a}. In
addition, a number of superdeformed bands of discrete $\gamma$-ray transitions were
discovered in selected $N = Z$ $\alpha$-like 
nuclei belonging to this mass region
(e.g.~\cite{Ide,Beck04}).
In this paper, we focus on the search for extended shapes and for the Jacobi
transition of $^{46}$Ti with
measurements of the LCP and GDR spectra in coincidence
with evaporation residues (ER) for the reaction $^{27}$Al~+~$^{19}$F at a bombarding
energy of $E_{lab}$($^{27}$Al)~=~144~MeV. In the following the
experimental results are presented and their 
interpretation are discussed within both the
statistical \cite{Pap04b,Pap03,Pul08} and the LSD \cite{Dub07} models.

\section{EXPERIMENTAL RESULTS AND INTERPRETATIONS}

The experiment was performed at the {\sc VIVITRON} tandem facility
of the IReS Strasbourg (France), using the multidetector array {\sc
ICARE} \cite{Pap04b,Bha09,Rou10} associated
with a large volume (4$^"\times$4$^"$) BGO detector.
 The $^{46}$Ti$^{*}$ compound nucleus (CN)
was populated
 at $E^*$~=~85 MeV and with $L_{max}$~$\approx$~35~$\hbar$
 by the inverse kinematics reaction 144~MeV $^{27}$Al on $^{19}$F.
High-energy $\gamma$-rays from the GDR decay were measured
using the BGO detector. The heavy fragments were detected in six
gas-silicon telescopes (IC), located at $\Theta_{lab} = \pm 10^{\circ}$ in three
distinct reaction planes. The in-plane detection of coincident LCP's was done
using ten triple telescopes (40~$\mu$m Si, 300~$\mu$m Si, 2~cm CsI(Tl) thick),
 eighteen two-element telescopes (40~$\mu$m Si, 2~cm CsI(Tl) thick), and four gas-silicon 
telescopes. Further details on the setup and on the calibration
procedures can be found in Ref. \cite{Brekiesz_Zakopane}.

\subsection{$\alpha$-particle spectra}

The energy spectra of the $\alpha$-particles emitted 
in the laboratory frame at the angles $\Theta_{lab}~=~45^{\circ}, 85^{\circ}, 125^{\circ}$
 in coincidence with $Z$ =18, 19, 20
measured by the IC located at $\Theta_{lab} =~10^{\circ}$ are
shown in Fig.~1 by the solid points.
The lines are the results of the analysis performed using {\sc CACARIZO}, the Monte
Carlo version of the statistical-model code {\sc CASCADE} ~\cite{Pul08}, which
is based on the Hauser-Feshbach formalism ~\cite{Bha09,Rou10,Mahboub}.
The high-energy part of $\alpha$-particle spectra depends 
on the final state level density. The
level density is calculated using the Rotating Liquid Drop Model (RLDM)
~\cite{Pul08} and can be changed using 
different sets of deformability parameters.
Larger deformations lower the yrast line and make it more flat, what increases the level density at
higher available excitation energy of the final nucleus, thus reduce $\alpha$-particle energies.
In the code, the yrast line is parameterized with deformability parameters 
$\delta_{1}$ and $\delta_{2}$: $E_{L} = \hbar^{2}L(L+1)/2\Im_{eff}$ 
with $\Im_{eff} = \Im_{sphere}(1+\delta_{1}L^{2}+\delta_{2}L^{4})$
~\cite{Bha09}, where $\Im_{eff}$ is the effective moment of 
inertia, $\Im_{sphere}$~=~(2/5)~$A^{5/3}r_{0}^{2}$
 is the rigid body moment of inertia 
of the spherical nucleus and $r_{0}$ is the radius parameter 
(set equal to $r_{0}$ = 1.3 fm in the present calculations).
 In Fig.~2 (left panel) the yrast lines used in the calculations are displayed as solid lines.
 The standard RLDM yrast line, which in general can be approximated by the rigid body yrast 
 line with small deformation ($\beta$ = 0.2), results in the $\alpha$-particle spectra presented in Fig.~1 
 as dashed lines. This parametrization does
not reproduce well the experimental spectra.

The solid line in Fig.~1 represents the calculations using the yrast line with 
the deformability parameters 
($\delta_{1}$~=~4.7$\times10^{-4}$, $\delta_{2}$~=~1.0$\times10^{-7}$)
proposed by \cite{Pap04b,Pap03}. 
This yrast line, labeled in Fig.~2 (left) 
"quasi-superdeformed"
as it resembles the yrast line for the rigid body with a deformation parameter 
$\beta$ = 0.6, reproduces 
fairly well the experimental spectra for $Z$ = 18 and $Z$ = 19. Such large deformations needed to explain the 
level density of the nuclei after evaporation of few particles are actually 
consistent with the superdeformed 
bands found in the nuclei with A$~\approx$~40~\cite{Ide,Beck04,Sven}.
\begin{figure}[htbp]
	\centering
		\includegraphics[scale=0.75]{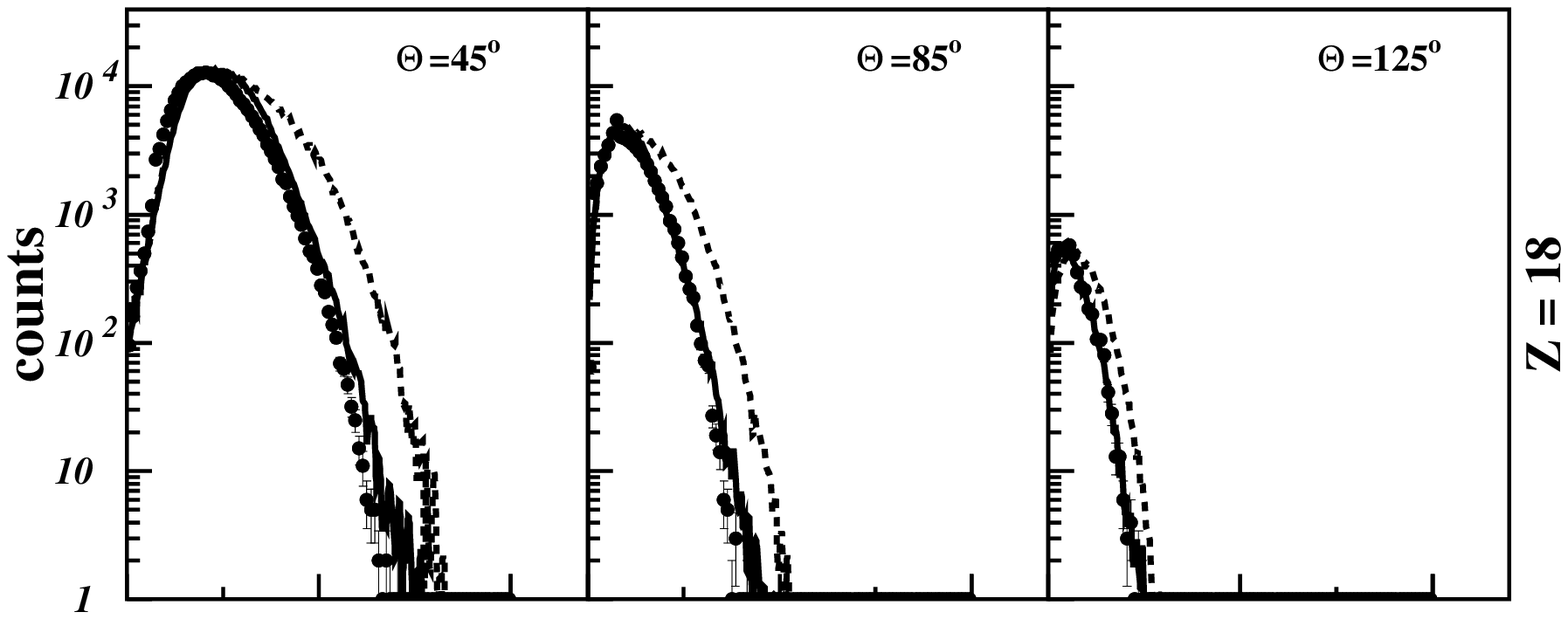}
		\includegraphics[scale=0.75]{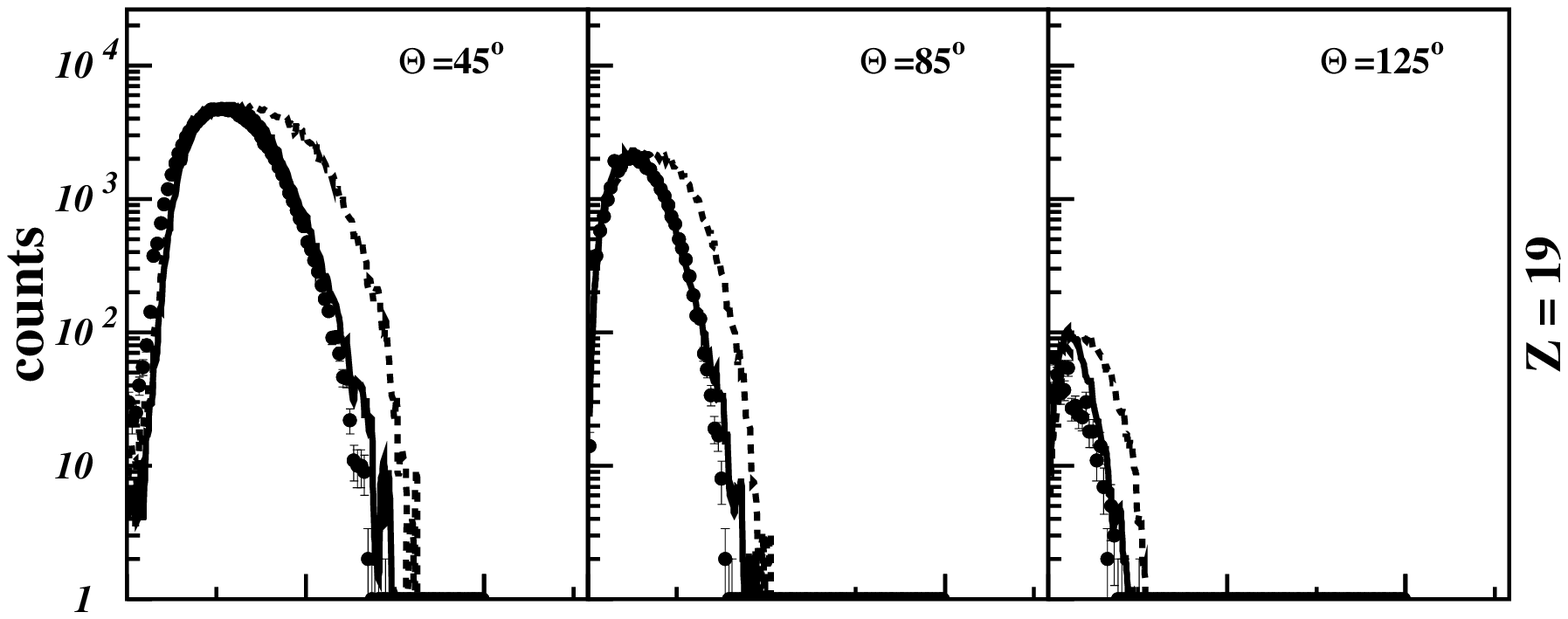}
		\includegraphics[scale=0.75]{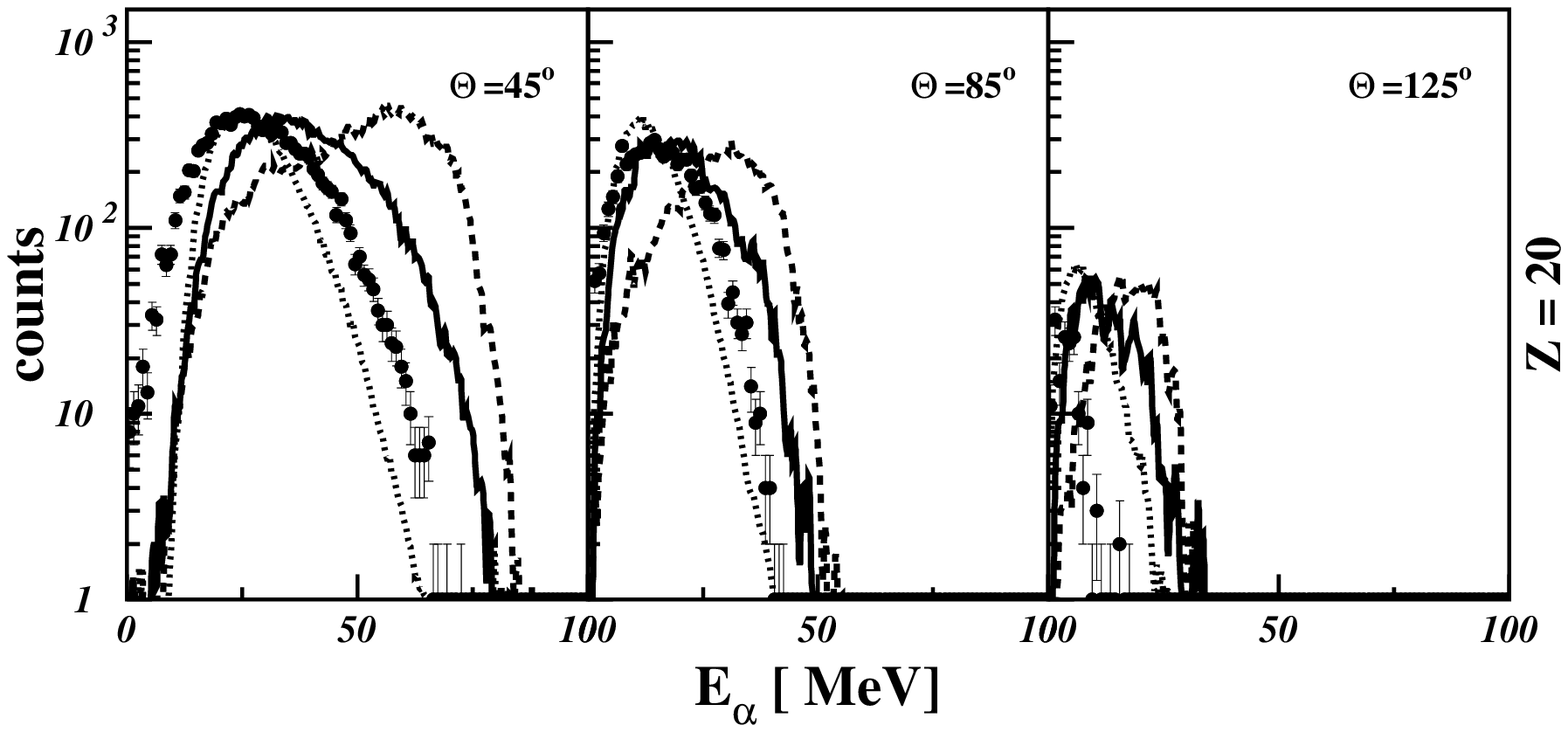}
		
	\caption{Experimental (points) and calculated (lines) $\alpha$-particle energy spectra for
$\Theta_{lab} = 45^{\circ},85^{\circ},125^{\circ}$
in coincidence with three different ER's ($Z$ = 18: upper row, $Z$ = 19: middle row,
$Z$ = 20: bottom row) detected in the IC at $\Theta_{lab}~=~10^{\circ}$. 
The calculations were carried out with yrast line deformation
parameters from RLDM (dashed line), for quasi-superdeformed shapes (solid
line) and for quasi-hyperdeformed shapes (dotted line).}
	\label{fig:Maj_fig1a}
\end{figure}
\begin{figure}[htbp]
	\centering
		\includegraphics[scale=0.9]{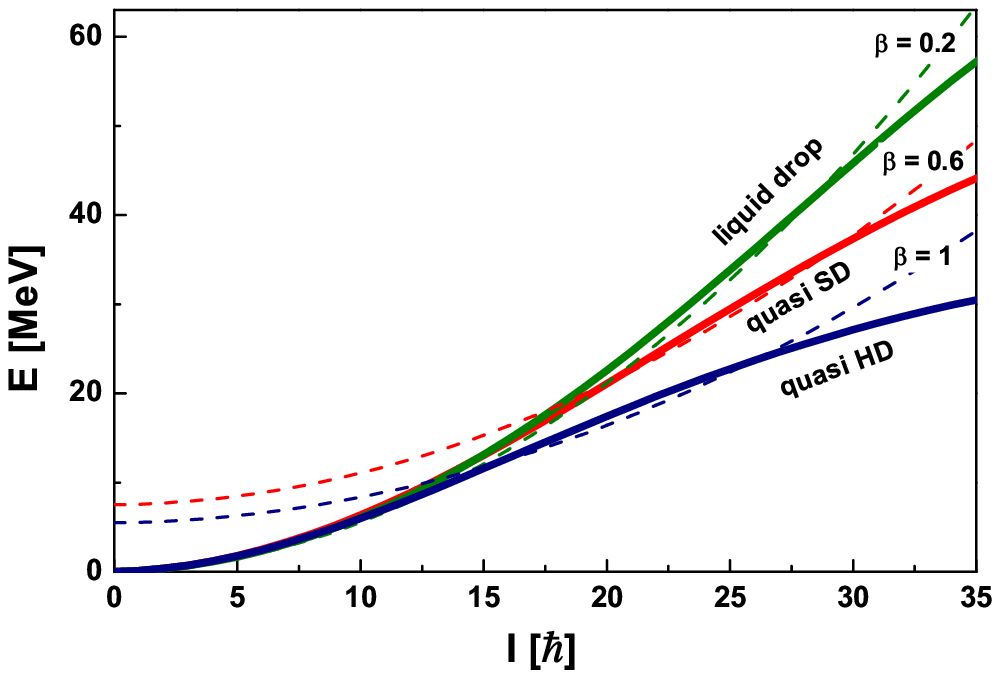}
		\includegraphics[scale=1.6]{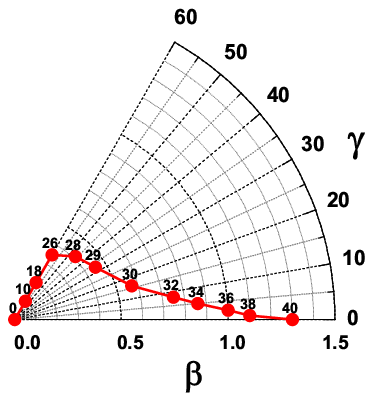}
	\label{fig:Maj_fig2}
	\caption{Left: the yrast lines used in the calculations (solid lines) 
and the rigid body yrast lines with different deformation parameters
 (dotted lines). Right: the evolution of the equilibrium shape of $^{46}$Ti
  as a function of spin predicted by the LSD model.}
\end{figure}

However, the spectra associated with $Z$ = 20, i.e. related to the emission of 
a single $\alpha$-particle,
are still in disagreement with the calculations using the RLDM or "quasi-superdeformed" yrast lines 
 (the slopes of the experimental spectra are larger),
 despite the fact that in $^{42}$Ca highly-deformed band was also found \cite{Lach},
  with the properties similar to the
 superdeformed bands in lower mass $\alpha$-like nuclei.
 In order to improve the agreement, an even more deformed yrast
line (which mimics an unusual extended shape) would be required.
 An example of the calculations in which the agreement of the slopes of the spectra is better,
 even the calculated slopes are too large, is shown
for $Z$ = 20 with dotted line.
 These calculations were using the yrast line with the deformability parameters 
($\delta_{1}$~=~1.1$\times10^{-3}$, $\delta_{2}$~=~1.0$\times10^{-7}$),
 depicted in Fig.~2 as "quasi-hyperdeformed" as it resembles the
 yrast line for the rigid body with a deformation parameter $\beta$~$\approx$~1.
This result seems to be rather unexpected.
Even though according to macroscopic-microscopic calculations made with
the LSD model, the equilibrium deformation of the (hot) nucleus $^{46}$Ti can
be characterized by such extreme deformations at the highest possible
spins of the reaction (I $\approx$ 34-36), the cooling process performed
through particle emission is expected to lead to much smaller
deformations. Indeed, at low temperatures shell effects play a prominent
role and tend to stabilize the nucleus in near-spherical or at best
super-deformed shapes \cite{Maj_Debrecen}. 
This is indeed the case when more than one charged particle is emitted 
(see the spectra for $Z$ = 18, 19 in Fig.~1).

A possible explanation can be related to the time scale of the evaporation 
process.
 When many particles are evaporated, the time needed for this process 
 can be long enough, so that the residual nucleus changes its shape
 to smaller deformation and the effective level density of final states can be described by the 
 "quasi-superdeformed" yrast line. In contrast, if only one $\alpha$-particle is being emitted (the case for $Z$ = 20),
 it is possible that the residual nucleus is still in the process of changing shape. Therefore the appropriate yrast
 line will have a shape between the shape of CN (with $\beta$~$\approx$~1),
 i.e. described by the "quasi-hyperdeformed" yrast line, and a shape of final nucleus, described by 
the "quasi-superdeformed" yrast line. Such an effect can be understood as an existence of a "dynamical hyperdeformation".
Although one cannot exclude another explanation of this anomalous shape of $\alpha$-particle spectra 
in coincidence with the residues of $Z$ = 20, e.g. spin dependent binding energies,
 it is worth to mention that analogous explanation was given for the observed $\alpha$-particle spectra 
in the decay of $^{59}$Cu CN \cite{Fornal}.

\subsection{GDR strength function from high-energy $\gamma$-ray spectrum}

The deformation parameters $\beta$ and $\gamma$ of the minima in the potential energy surfaces, calculated with the LSD model,
 are presented in the Fig.~2 (right panel), showing the evolution of the equilibrium shape of $^{46}$Ti as function of spin.
 As angular momentum increases, the nucleus originally spherical
 at I = 0 acquires some oblate deformation, corresponding to an elongation of up to
$\beta$~$\approx$~0.3 for I = 26. Beyond 26 $\hbar$, the Jacobi shape transition sets in:
 the nucleus becomes triaxial with the elongation parameter increasing up to values of $\beta$~$\geq$~1 for I = 38,
 and the fission finally takes place at I~$\approx$~40.
\begin{figure}[htbp]
	\centering
		\includegraphics[scale=1.2]{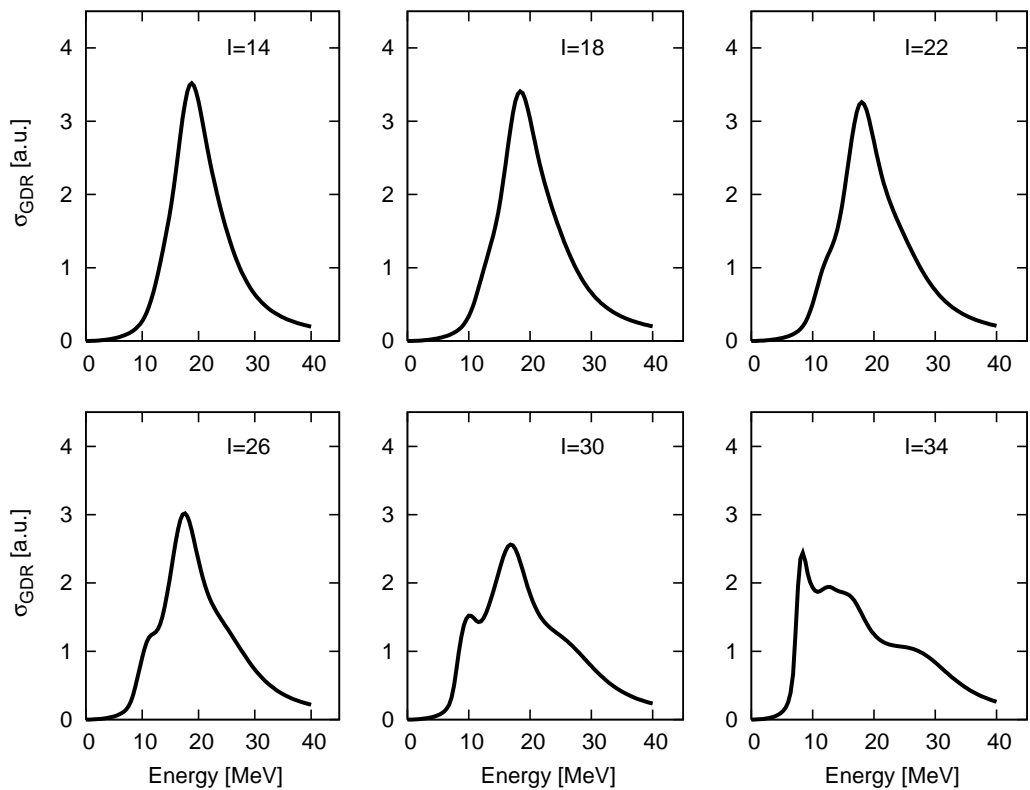}
	\label{fig:Maj_fig3}
	\caption{GDR strength functions for $^{46}$Ti at different spins simulated by the thermal shape fluctuation model
	 based on the potential energies from the LSD model.}
\end{figure}

 The LSD potential energy surfaces constituted also a basis
 for the thermal shape fluctuation method to compute the GDR strength functions \cite{Dub07}, presented in Fig.~3.
 For the oblate regime below I = 26 the calculated GDR line shape has the form of a broad Lorentzian distribution. 
 For the Jacobi regime (I~$\geq$~26) the line shape is characteristically 
split forming a narrow low-energy component
  around 10 MeV, and a broad structure ranging from 15 to 27 MeV. This splitting is a consequence of both the 
 elongated shape of the nucleus undergoing the Jacobi shape transition and strong Coriollis effects \cite{Nergard} 
 which additionally split the low energy component and shift a part of its strength down to 10~MeV. Such fragmented GDR strength
 function was indeed experimentally observed for $^{46}$Ti at highest spins in the EUROBALL+HECTOR experiment \cite{Maj02}.
\begin{figure}[htbp]
	\centering
	\includegraphics*[scale=1.6]{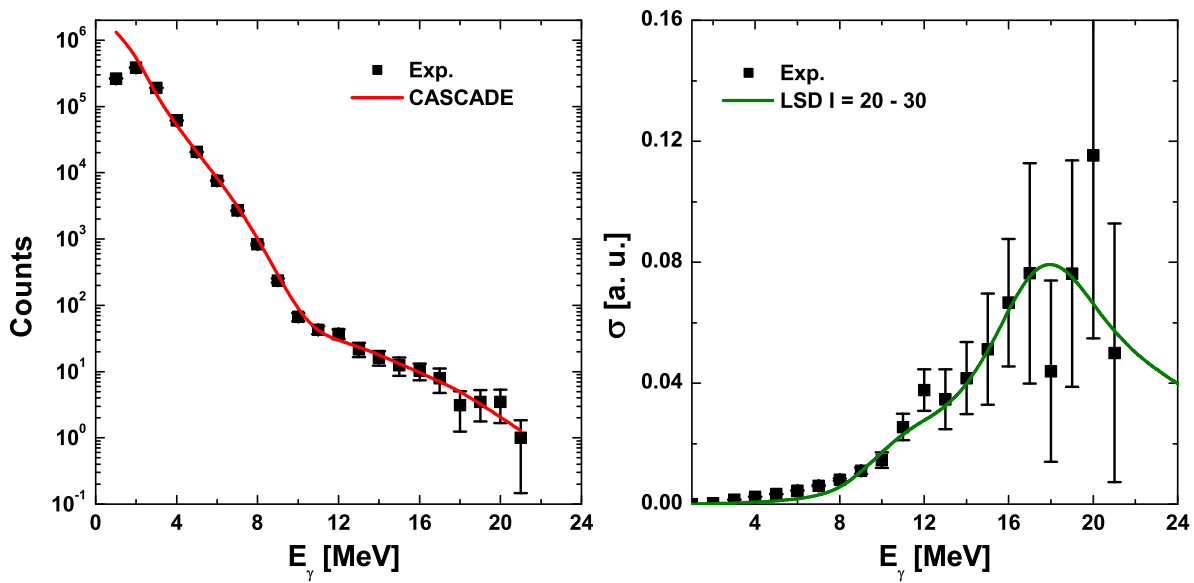}
	\label{fig:Maj_fig4}
	\caption{Left: High-energy $\gamma$-ray spectrum measured in coincidence with the recoils possessing $Z$ = 18, 19 and 20,
	 together with the best fit CASCADE calculations in which the "quasi-superdeformed" yrast line was used and the GDR 
	 parameters that fit the spectrum were $E_{GDR}$ = 18.5 MeV, $\Gamma_{GDR}$ = 14 MeV and $\sigma_{GDR}$ = 1.
	 Right: The extracted GDR strength function (points) together with the LSD model prediction (line) for I~$\geq$~20.}
	\end{figure}

In the present experiment there was only very limited way to select the GDR 
decay related to the highest spins,
 namely by selecting the coincidences with residues possessing highest $Z$ = 20. This was sufficient 
 for the $\alpha$-particle channel selection
although the statistics of the GDR $\gamma$-decay was very low.
 Therefore in order to perform the statistical model analysis the sum of coincidences with $Z$ ranging from 18 to 20 was used.
 The resulting high energy $\gamma$-ray spectrum is shown in Fig.~4 (left panel), together with the CASCADE 
 calculations in which the GDR parameters were fitted. The extracted GDR strength function is shown
in the right panel of Fig.~4. Indeed the data do not show any distinct low energy component, but this is consistent 
with the LSD model prediction, when one takes into account the fact that this spectrum 
consists of contributions from a broad spin window
($\geq$~20 $\hbar$).

\section{SUMMARY AND OUTLOOK}

The spectra of $\alpha$-particles emitted from hot rotating $^{46}$Ti 
evidence the large deformations 
($\beta$~$\approx$~0.6) that are involved in the evaporation process.
The process of emission of a single $\alpha$-particle suggests even larger 
deformations linked to the effect of "dynamical hyperdeformation". The GDR 
spectra are also consistent with such large deformations. Unfortunately,
the lack of the effective high-spin selectivity in the experiment does not 
allow us to observe the predicted GDR splitting. However, the global
properties of the GDR spectrum of Fig.~4 are fairly well reproduced 
by the LSD model.

This type of experiments will be attractive with the availability of the intense radioactive beams, 
 since in the more neutron rich nuclei, produced in fusion-evaporation reactions,
 higher spins can be reached, enabling stronger population of the "Jacobi shapes" window.

This work has been supported by the Polish Ministry of Science and Higher
Education (Grant No. 1 P03B 030 30). We thank the VIVITRON staff, J. Devin
and M.A. Saettel for the excellent support in carrying out the experiment.

\end{document}